\documentclass[usenatbib,conference]{basi}
\usepackage[british]{babel}
\usepackage[varg]{txfonts}
\usepackage{rotating}
\usepackage{dcolumn}
\begin{document}
\title{Prominence Formation and Oscillations}
\author[P.~F.~Chen]%
       {P.~F.~Chen$^{1,2}$\thanks{email: \texttt{chenpf@nju.edu.cn}} \\
       $^1$Department of Astronomy, Nanjing University, Nanjing 210093,
        China\\
       $^2$Key Lab of Modern Astron. \& Astrophys. (Ministry of
        Education), Nanjing University, China}

\pubyear{2013}
\volume{10}
%\pagerange{\pageref{1}--\pageref{10}}
\pagerange{\pageref{1}--10}

\date{Received \today}

\maketitle
%------------------------------------------------------------------------------%
% abstract and keywords                                                        %
%------------------------------------------------------------------------------%
\label{1}

\begin{abstract}
Prominences, or filaments, are a striking phenomenon in the solar atmosphere.
Besides their own rich features and dynamics, they are related to many other
activities, such as solar flares and coronal mass ejections (CMEs). In the past
several years we have been investigating the prominence formation,
oscillations, and eruptions through both data analysis and radiative
hydrodynamic and magnetohydrodynamic (MHD) simulations. This paper reviews our
progress on these topics, which includes: (1) With updated radiative cooling
function, the coronal condensation becomes a little faster than previous work;
(2) Once a seed condensation is formed, it can grow via siphon flow
spontaneously even if the evaporation stops; (3) A scaling law was obtained to
relate the length of the prominence thread to various parameters, indicating
that higher prominences tend to have shorter threads, which is consistent with
the fact that threads are long in active region prominences and short in 
quiescent prominences; (4) It was proposed that long-time prominence
oscillations out of phase might serve as a precursor for prominence eruptions
and CMEs; (5) An ensemble of oscillating prominence threads may explain the
counter-streaming motion.
\end{abstract}


%------------------------------------------------------------------------------%
% main text of the paper, using \section, \subsection, \subsubsection          %
%------------------------------------------------------------------------------%
\section{Introduction}\label{s:intro}

Filaments are striking features in the solar atmosphere, typically observed in
chromospheric lines like H$\alpha$. They are also identifiable in EUV images.
Against the solar disk, they appear as long filamentary structures, by which
the phenomenon was called. When they appear above the solar limb, it was
recognized that they are dense plasma suspended in the hot tenuous corona. In
this case they are called prominences. The two terminologies are used interchangeably in the literature. With the temperature $\sim$100 times lower
and the density $\sim$100 times higher than the ambient corona, prominences are
interesting to researchers in various aspects. Their formation and dynamics are
related to the energy transport (heating and cooling) and force balance 
\citep{low12} in the corona; Their oscillations can be applied to diagnose
the magnetic field where the prominence is embedded; Their eruptions are then
directly related to solar flares and coronal mass ejections (CMEs). It has been
established that the erupting prominence is the core of the CME \citep{hous81}.
Here we emphasize that prominences might also be the core of CME researches.
This is reflected in the various aspects of the CME-related researches. First,
prominences are formed in highly sheared magnetic field, including flux ropes
and sheared arcades, where the long flux tubes favor thermal instability by
which prominences are often believed to be formed. Highly sheared magnetic 
field is also a signature of the nonpotentiality of the magnetic system which
possesses sufficient magnetic free energy to power CMEs. Second, the initiation
of the filament might well correspond to the onset of CMEs, when the CME frontal
loop has not yet formed \citep{chen11}. In particular, the mass drainage from
the prominence might serve as one possible triggering mechanism for CMEs.
Third, prominences are the core of the CMEs and the magnetic cloud of 
interplanetary CMEs. Fourth, regarding the debate whether the flux rope is
formed before or during CME eruptions \citep{zhan12a, cheng13}, \citet{chen11}
suggested that in many cases a flux rope exists in the progenitor of the CME,
which corresponds to the eruption of an inverse-polarity prominence, whereas in
other cases the flux rope is formed during CME eruption via magnetic
reconnection, which corresponds to the eruption of a normal-polarity prominence.

There are several detailed review papers on prominences \citep{mart98, labr10,
mack10, schm12}. In this paper, we summarize what my group have done on
prominences in recent years, with the purpose to clarify what are the best
follow-up researches in the foreseeable future. The topics cover the formation, 
dynamics, oscillations, and eruptions of prominences.

\section{Formation mechanism}\label{s:form}

The formation mechanisms of cold prominence plasma remain to be a controversial
issue. As reviewed by \citet{mack10}, there are basically two mechanisms, i.e.,
direct injection model (from the chromosphere to the corona) and the 
evaporation-condensation model. The latter came from the idea of thermal
instability of the coronal plasma \citep{park53}. Noticing that the coronal
plasma along the flux tube is not sufficient to supply the necessary mass for
the prominence thread, it has been proposed that there is chromospheric 
evaporation to the corona before condensation, which is driven by extra heating
localized in the chromosphere \citep[e.g.,][]{pola86}. Such an 
evaporation-condensation model was numerically simulated by \citet{mok90},
\citet{anti99}, and \citet[][and references therein]{karp08}, and was 
demonstrated to be able to explain various observational features. Note that
all these simulations were done in one dimension (1D). This is validated since 
high-resolution observations indicate that prominences are composed of many
narrow threads which are supposed to run along the individual magnetic flux
tubes \citep{lin08}, and threads are the building blocks for prominences.

\begin{figure}
\centerline{\includegraphics[width=8.5cm]{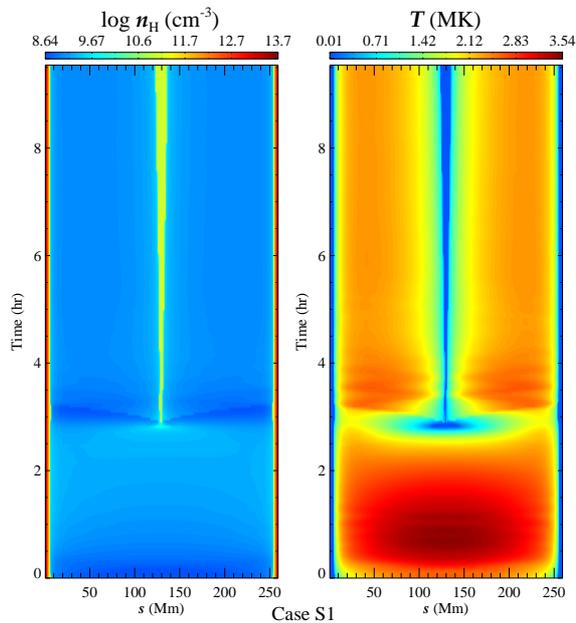}}
\caption{Temporal evolution of the plasma density (left) and the temperature
      (right) along the model loop in the case of symmetric heating. The two
      loop footpoints are at s = 0 and 260 Mm, respectively 
	\citep[from][]{xia11}.
        \label{fig1}}
\end{figure}

Following this line of thought, we adopted the recently updated radiative
cooling function to numerically simulate the response of the solar atmosphere
to the enhanced chromospheric heating by solving 1D radiative hydrodynamic
equations \citep{xia11}. Figure \ref{fig1} depicts the evolution of the density
and temperature distributions along the magnetic loop in the case of symmetric
heating. It is seen that with the heating rate $E_1=10^{-2}$ erg cm$^{-3}$
 s$^{-1}$ in the chromosphere, the coronal condensation occurs 2.8 hr after the
extra heating is introduced to the chromosphere. In contrast, the radiative
cooling function used in \citet{karp05} was 1-2 times lower than ours, and the
coronal condensation occurs at $t=$3.5 hr. 

%---------------------------------------------------------------------%
%--------------------------------------------------------------%

%---------------------------------------------------------------------%
\begin{figure}
\centerline{\includegraphics[width=10cm]{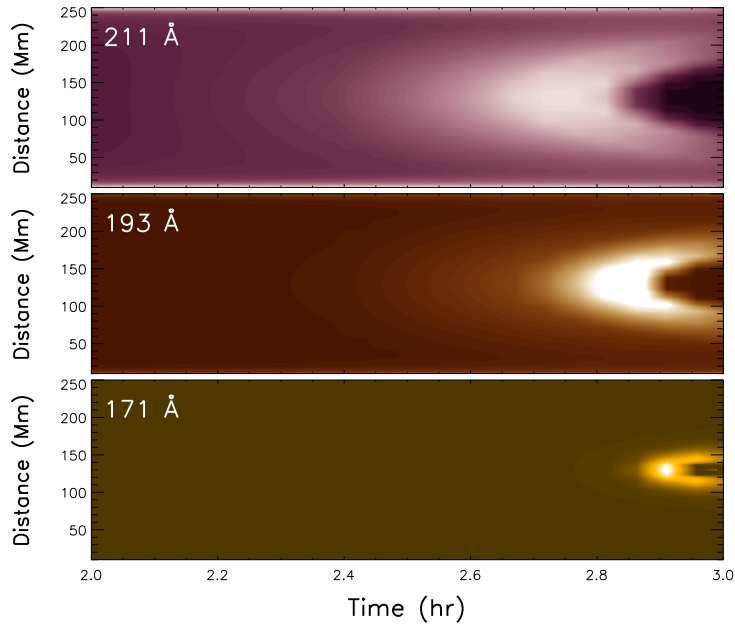}}
\caption{Temporal evolutions of EUV intensity distributions along the model
loop at three wave bands, Fe {\small XIV} 211 \AA, Fe {\small XII} 193 \AA,
and Fe {\small IX} 171 \AA. The EUV intensity is calculated from the temperature and density distributions in Fig. \ref{fig1}.      \label{fig2}}
\end{figure}
%--------------------------------------------------------------%

The evaporation-condensation model was recently directly confirmed by 
\citet{liu12} and \citet{berg12} with the EUV observations from {\it Solar
Dynamics Observatory} ({\it SDO}), which showed that the enhanced emissions
appeared in turn from higher-temperature wave band (211 \AA) to
lower-temperature wave band (171 \AA) and all the way to He {\small I} 304
{\AA} which is mainly from the cool 
prominence plasma. In order to compare our simulation results with the 
observations, we calculate the EUV intensity distribution along the magnetic
loop, whose temporal evolutions at three wave bands are presented in Fig.
\ref{fig2}. The numerical results are seen to be qualitatively consistent with
observations very well, i.e., just before the prominence formation the strong
EUV emission appears in the hot line (Fe {\small XIV} 211 \AA) first, and then
shifts to the lower-temperature lines successively. However, one big difference
between our Fig. \ref{fig2} and the Figure 3 of \citet{berg12} is that the time
delay of the strong emissions between 171 {\AA} and 211 {\AA} is $\sim$12 hr in
\citet{berg12}, whereas it is 12 min in our Fig. \ref{fig2}, i.e., 60 times
shorter than in the observations. If the radiative cooling function is precise,
there are two other possible reasons for the discrepancy. One is that the
background heating rate might be under-estimated. The other is that the plasma
density just prior to the coronal condensation in \citet{berg12} is many
times smaller than the value in our simulations, i.e., $n=2\times10^9$ 
cm$^{-3}$, considering the radiative cooling timescale is proportional to
$1/n$. In this case, if we want to explain the observations in \citet{berg12},
the extra chromospheric heating rate used for Fig. \ref{fig1}, i.e., 
$E_1=10^{-2}$ erg cm$^{-3}$ s$^{-1}$, might be too large so that the strong
evaporating flow compresses the coronal plasma at the loop apex to a high
value. This issue should be investigated in the future. As far as \citet{xia11}
can tell, the decreasing $E_1$ does postpone the formation of prominences, and
also increases the delay of EUV emissions from hot to warm wave bands.

\section{Length and growth rate}

Observations indicate that the threads of active region prominences are often
long and the threads of quiescent prominences are short \citep{mack10}. This is
an interesting property which deserves a sound explanation. Apparently it may 
be directly related to the difference of the magnetic environment between the
two types of prominences since active region prominences are generally low and
quiescent prominences are located several times higher. Recently, we did a
parameter survey on how the magnetic configuration influences the features of
the prominences \citep{zhan13}. It is revealed that if other factors, e.g., the
width and the depth of the magnetic dip, and the evaporation time, are kept the
same, the length of the prominence thread ($l$) scales with the altitude ($h$)
by $l\sim h^{-0.37}$, which implies that higher prominences tend to have
shorter threads, which is consistent with observations.

%---------------------------------------------------------------------%
\begin{figure}
\centerline{\includegraphics[width=6.2cm]{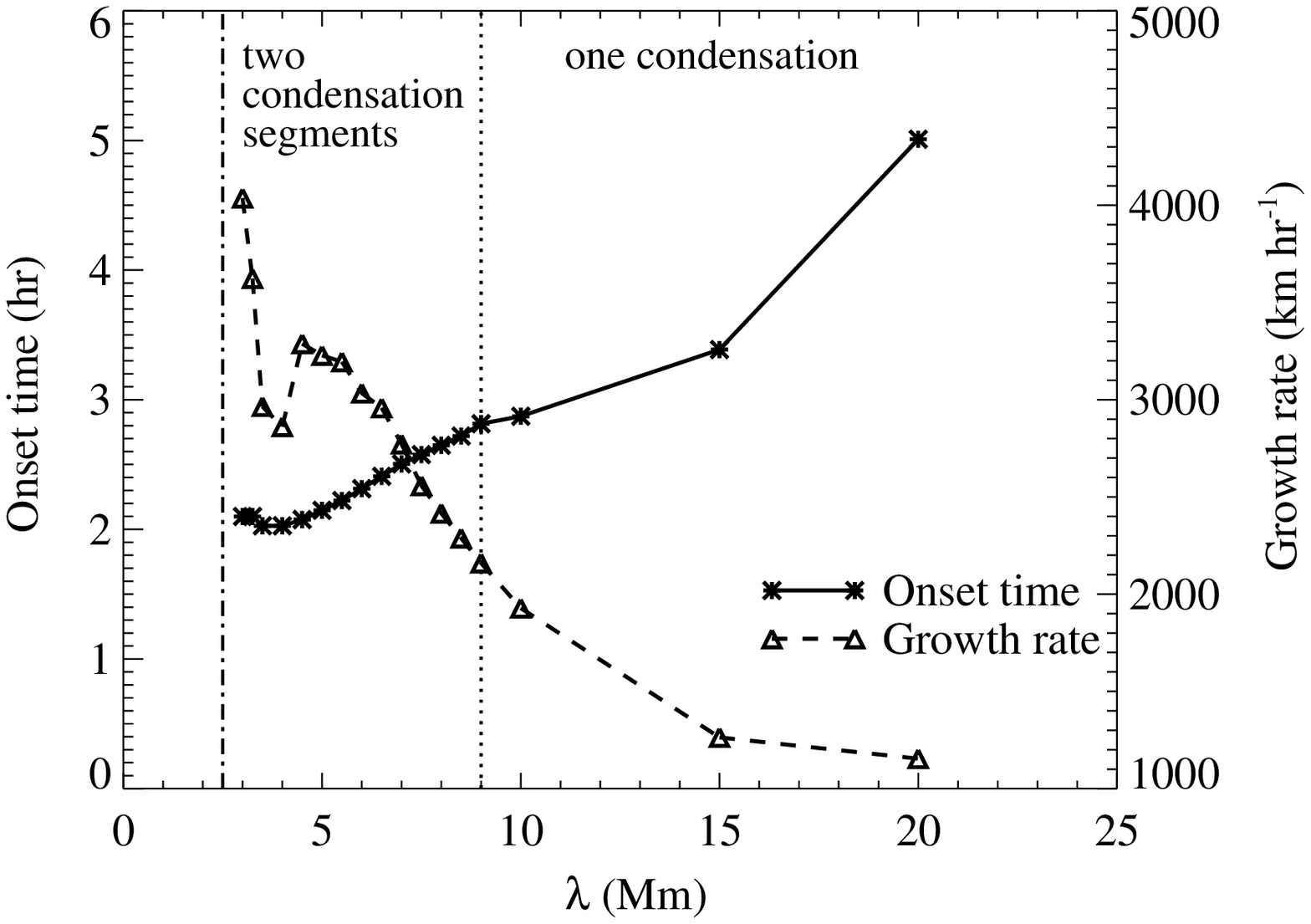}\hspace{8pt}
\includegraphics[width=6.2cm]{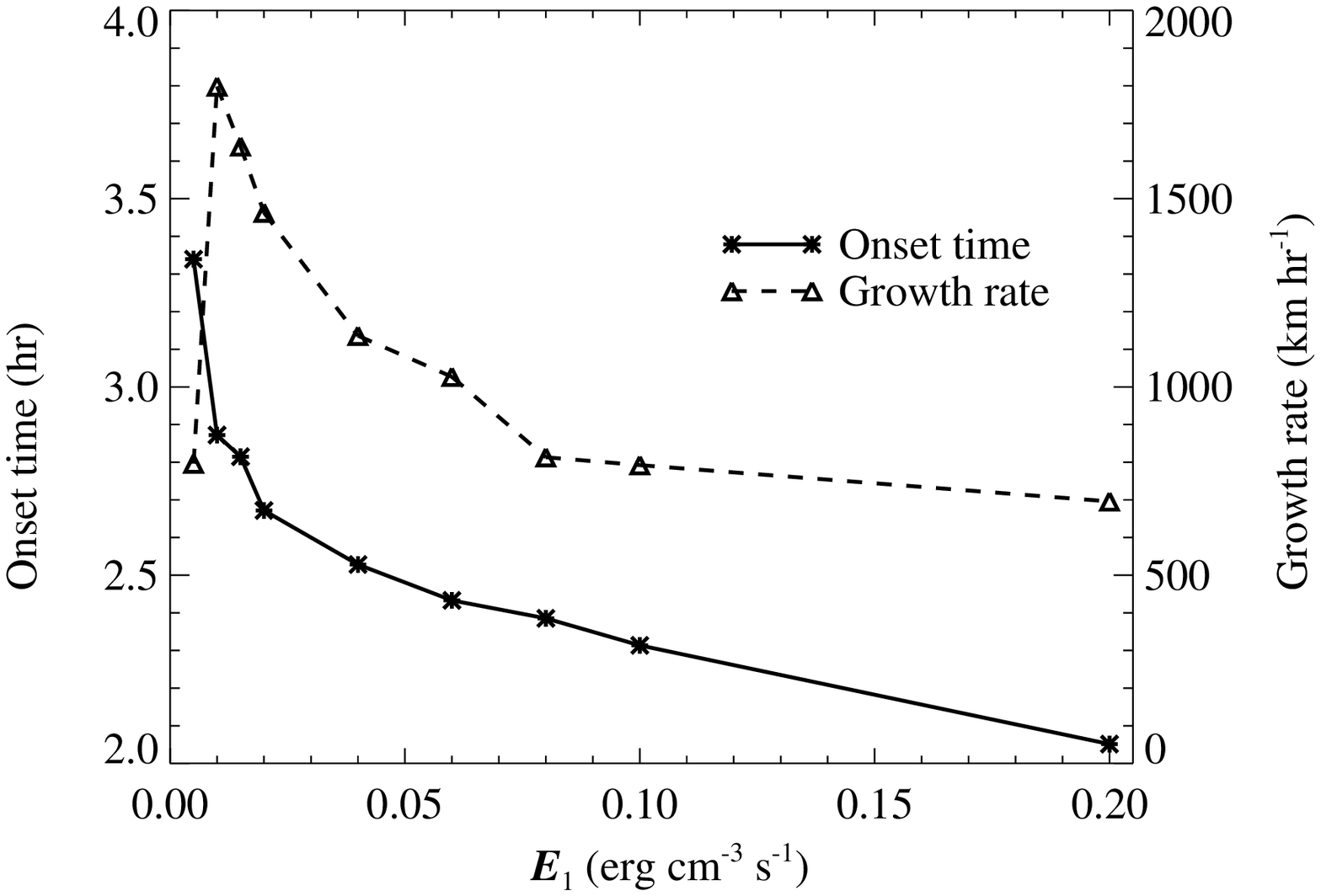}
}
\caption{The dependence of the condensation onset time (asterisks) and the
growth rate (triangles) on the amplitude ($E_1$) and the scale height
($\lambda$) of the extra chromospheric heating  \citep[from][]{xia11}.
        \label{fig3}}
\end{figure}
%--------------------------------------------------------------%

The long thread is not formed at once. It grows from seed condensation in the
corona. Various parameters may affect the growth rate of the prominence thread.
\citet{xia11} found that both the amplitude ($E_1$) and the scale height 
($\lambda$) of the extra chromospheric heating change the growth rate in a
non-monotonic way. As shown by the triangles in Fig. \ref{fig3}, both weak and
strong $E_1$ do not favor the fast growth of the prominence thread, whereas the
growth rate decreases with increasing $\lambda$ except a significant drop near 
$\lambda$=3.5 Mm.

\citet{xia11} noted that once radiative instability happens near the loop apex
due to chromospheric evaporation, the tradeoff of the decreasing temperature 
and the increasing density turns out that the gas pressure decreases to a value
smaller than in the pre-evaporation stage. As a result, even if the extra
chromospheric heating is switched off, a siphon flow is formed naturally due to
the loss of force balance which enables the initial static coronal loop:
chromospheric plasma is siphoned up and then heated in the corona via 
background heating. When the siphon flow penetrates into the cold condensation
near the loop apex, it is cooled via enhanced radiation. This implies that
extra chromospheric heating is not necessary to maintain or grow the prominence
thread. Once a seed condensation is formed, it would grow without the help of
extra chromospheric heating and evaporation, although the growth rate is
smaller than in the case with extra chromospheric heating. This property might
be meaningful since the extra heating, which is responsible for the
chromospheric evaporation, might be due to low-atmospheric magnetic
reconnection \citep[e.g.,][]{chen01}, which generally has a limited lifetime.
It might be not realistic to have continual heating.

A further question arises following the above-mentioned simulations, that is,
given a dipped magnetic loop, what is the maximum length that the prominence
thread can grow. This question is being tackled currently \citep{zhou14}.

\section{Oscillations}

The solar corona is always dynamic, and disturbances are ubiquitous, e.g.,
from sporadic CMEs, flares or subflares, to the non-stopping convection flows
in the photosphere, which drive kink/Alfv\'en waves into the corona 
\citep{tian12}. Therefore, once a prominence is formed it is subjected to
all these disturbances, and is ready to oscillate. Prominence oscillations can
be divided into large-amplitude versus small-amplitude ones, or transverse 
versus longitudinal ones \citep{arre12}. The observational characteristics,
including the period and damping timescale, can be used to diagnose the thermal
and magnetic parameters of the prominences. Since longitudinal oscillations
can be simulated in 1D, we investigated this topic as a start for its
simplicity.

%---------------------------------------------------------------------%
\begin{figure}
\centerline{\includegraphics[width=7.5cm]{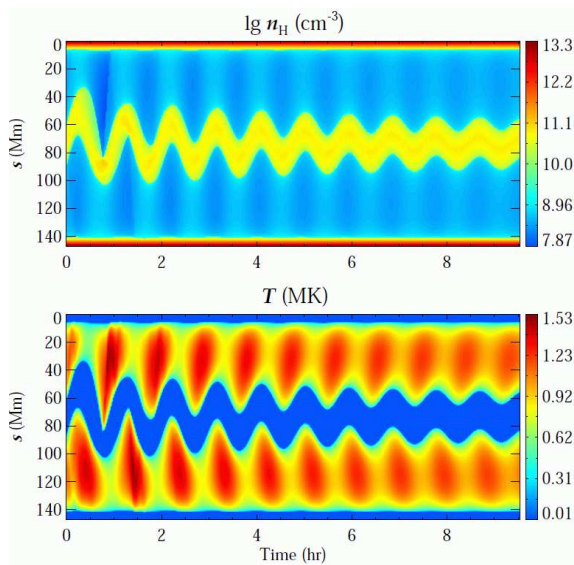}}
\caption{Time evolutions of the density ({\it top}) and the temperature
({\it bottom}) distributions along the magnetic loop, which indicate that
the prominence experiences a damped oscillation subjected to a perturbation 
   \citep[from][]{zhan12b}.
        \label{fig4}}
\end{figure}
%--------------------------------------------------------------%

With the magnetic geometry derived from observations, we \citep{zhan12b} 
numerically simulated the prominence oscillations, as depicted in Fig. 
\ref{fig4}, which shows the temporal evolutions of the density and the
temperature distributions along the magnetic loop. We found that the simulations can reproduce the period
of the longitudinal prominence oscillation which was observed on 2007 February
8, though the resulting damping timescale is 1.5 times longer than the
observational value.

Recently, we \citep{zhan13} did a parameter survey about the characteristics of
prominence oscillations. We first compared the effect of the trigger type, 
i.e., localized heating or impulsive momentum from a nearby subflare., which
turned out to be that the oscillation is nearly independent of the trigger 
type. It was found that with the presence of non-adiabatic terms including
thermal conduction and radiative cooling the oscillation would damp out, where 
the radiative cooling was demonstrated to be dominant. Scaling laws were
obtained to relate the oscillation period ($P$) and decay timescale ($\tau$) to
various parameters, i.e., $P\sim 2\pi\sqrt{R/g_{\odot}}$, where $R$ is the 
curvature radius of the magnetic dip, and $\tau\sim l^{1.63}D^{0.66}w^{-1.21}
v_{0}^{-0.30}$, where $l$ is the prominence length, $D$ and $w$ are the depth
and the width of the magnetic dip, and $v_0$ the velocity perturbation
amplitude. The scaling law for $P$, which is the same for a pendulum, implies
that the field-aligned component of the gravity is the main restoring force for
the longitudinal oscillations, as also found by \citet{luna12}.

Besides the scaling laws, two more results are worth mentioning. One is that
we found that if a subflare occurs immediately near the footpoints of the 
magnetic loops running through the prominence, $\sim$4\% of the released thermal
energy would be converted to the kinetic energy of the prominence oscillation.
The other one is that we found that mass drainage from the prominence to the
chromosphere would significantly damp the oscillation.

\section{Long-time oscillation as a precursor for CMEs}

\begin{figure}
\centerline{\includegraphics[width=11cm]{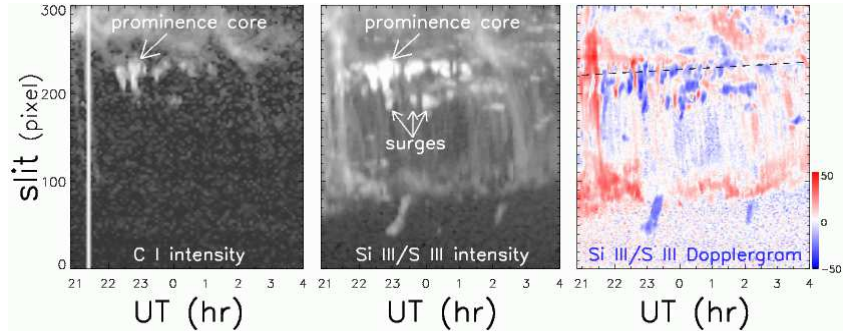}}
\caption{Left panel: Evolution of the C {\small I} 1118.45 {\AA}
      intensity along the SUMER slit; Middle panel: Same for
      S {\small III}/Si {\small III} 1113 {\AA}; Right panel: Evolution of
      the Dopplergram along the the SUMER slit observed at
      S {\small III}/Si {\small III} 1113 {\AA} 
   \citep[from][]{chen08}.
        \label{fig5}}
\end{figure}

Generally prominence oscillations damp in $\sim$3--4 periods, as found both
in observations and simulations mentioned above. However, sometimes the
prominence oscillation may persist for much longer time. With SUMER 
spectrometer, we presented a case where a prominence oscillated for 12 periods
before eruption \citep{chen08}, as displayed in Fig. \ref{fig5}. With that, we
proposed that long-time prominence oscillation may be a precursor for 
prominence eruptions and CMEs.

%---------------------------------------------------------------------%
%--------------------------------------------------------------%

Such a proposal was backed by recent two examples, and in both events the
prominence was oscillating longitudinally before eruption. With {\it Hinode}
and {\it SOHO} observations, \citet{zhan12b} analyzed an event where only a
thread of a prominence erupted to form a CME, with the main body remaining at
the original height. They found that before the thread eruption, the whole
prominence body was oscillating along its spine. With high-resolution 
observations from {\it SDO}, \citet{li12} revealed the longitudinal oscillation
and the ensuing eruption of the whole filament, where plasma drainage from the
oscillating filament to the solar surface may facilitate the onset of the
eruption.

Such prominence oscillations would continue in the later eruption phase as
revealed by \citet{isob06} and \citet{mier12}.

\section{Counter-streaming}

Counter-streaming of plasma was found in prominences \citep{zirk98}. Its nature
is still unclear, and we are still not sure whether it is common in any
prominence at any time or it is a signature of prominence activation.

%---------------------------------------------------------------------%
\begin{figure}
\centerline{\includegraphics[width=8cm]{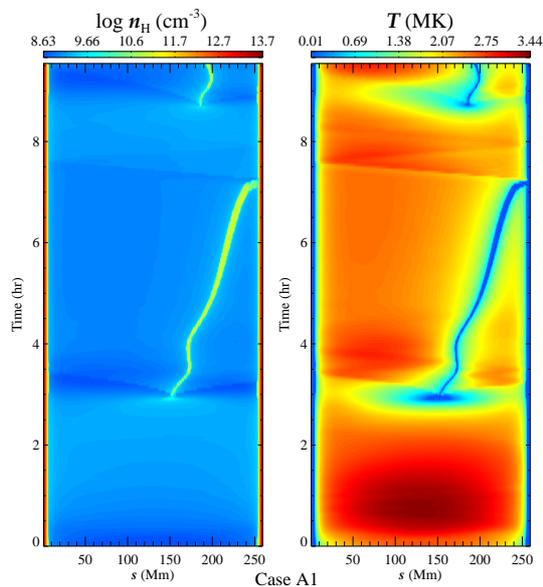}}
\caption{Temporal evolution of the plasma density ({\it left}) and the
   temperature ({\it right}) along the model loop in the asymmetric. The loop
   footpoints are at $s=0$ and $260$ Mm, respectively 
   \citep[from][]{xia11}.
        \label{fig6}}
\end{figure}
%--------------------------------------------------------------%

The first possibility is favored by the longitudinal prominence oscillations.
The longitudinal oscillations can be extrinsic, i.e., being triggered by a 
nearby subflare, or intrinsic, i.e., through asymmetric heating at the two
footpoints of the magnetic loop. Similar to \citet{anti99} and \citet{luna12},
we \citep{xia11} found that if the extra chromospheric heating is asymmetric
at the two footpoints of the magnetic loop, the prominence, upon formation,
would oscillate around the magnetic dip or even flow along the loop to drain
down toward the footpoint with weaker heating and then repeat the 
formation-drainage cycle, as illustrated by Fig. \ref{fig6}. The prominence
thread experiences oscillations while moving to the right. Counter-streaming
might just be an ensemble of oscillating threads which are not in phase 
\citep[a similar idea was mentioned by][]{ahn10}. In practice, even if the
oscillations of different threads are in phase initially, they would evolve to
be out of phase since different threads have different oscillation periods.
Furthermore, even in the case without longitudinal oscillations, 
counter-streamings may still appear if the footpoint heating is randomly
distributed in the solar surface, which drives mass drainage in a random way,
i.e., the drainage is toward the positive magnetic polarity in some threads
and toward the negative polarity in others. With these possibilities, we
speculate that counter-streaming may not necessarily be the precursor for
prominence eruptions and CMEs. However, the significant change of the
counter-streaming might serve as a precursor for prominence eruptions and CMEs,
which should be clarified in the future. Our recent results suggest that both
longitudinal and interlaced uni-directional flows contribute to
counter-streamings \citep{chen14}.

\section{Prospects}

Aided by the high-resolution multi-wavelength observations from various
telescopes, more and more detailed features of prominences and their dynamics
are being revealed, which provide evidence for the evaporation-condensation
model and open new windows for theoretical \citep{low12} and numerical studies
\citep{xia11, luna12}. Such a model, which was numerically realized only in 1D
radiative hydrodynamics until 2011, was extended to 2.5D MHD by \citet{xia12},
while its extension to 3D MHD is also on-going, which will be
crucial to the understanding of the detailed observations of prominences.

\section*{Acknowledgements}

PFC is grateful to the SOC members for the invitation to present the review
paper and to my students for their contributions. This work was financially
supported by Chinese foundations 2011CB811402 and NSFC (11025314, 10878002,
and 10933003).

%------------------------------------------------------------------------------%

\label{lastpage}
%------------------------------------------------------------------------------%
\end{document}